# Modified Bully Algorithm using Election Commission

Muhammad Mahbubur Rahman, Afroza Nahar

*Abstract* — Electing leader is a vital issue not only in distributed computing but also in communication network [1, 2, 3, 4, 5], centralized mutual exclusion algorithm [6, 7], centralized control IPC, etc. A leader is required to make synchronization between different processes. And different election algorithms are used to elect a coordinator among the available processes in the system such a way that there will be only one coordinator at any time. Bully election algorithm is one of the classical and well-known approaches in coordinator election process. This paper will present a modified version of bully election algorithm using a new concept called election commission. This approach will not only reduce redundant elections but also minimize total number of elections and hence it will minimize message passing, network traffic, and complexity of the existing system.

*Index Terms*— Bully election algorithm, coordinator, election Commission, message passing,

## I. INTRODUCTION

IN a distributed computing system, a process is used to coordinate many tasks. It is not an issue which process is doing the task, but there must be a coordinator that will work at any time. So electing a coordinator or a leader is very fundamental issue in distributed computing. And there are many algorithms that are used in election process. Bully election algorithm is one of them. This paper represents a modified version of bully algorithm using a new concept Election Commission. It reduces redundant elections, minimizes message passing and network traffic. In section 2, it is given a very short idea about election algorithm, section 3 represents original bully algorithm, different modified version of bully algorithms, their procedures and limitations. Methodology of our proposed algorithm is given in section 4. Section 5 describes our proposed algorithm with proper examples. In Section 6, an overall comparison of our algorithm with bully algorithm and existing modified bully algorithms is given.

Muhammad Mahbubur Rahman is with the department of Computer Science in American International University-Bangladesh, (corresponding author to provide phone: +8801712721654; e-mail: mahbubr@aiub.edu)

Afroza Nahar is with the department of Computer Science in American International University-Bangladesh, (e-mail: afroza@aiub.edu)

## II. ELECTION ALGORITHM

An election algorithm is an algorithm for solving the coordinator election problem. Various algorithms require a set of peer processes to elect a leader or a coordinator. It can be necessary to determine a new leader if the current one fails to respond. Provided that all processes have a unique identification number, leader election can be reduced to finding the non crashed process with the highest identifier.

## III. BULLY ALGORITHM AND ITS DIFFERENT UPDATES

An election algorithm is an algorithm for solving the coordinator election problem. Various algorithms require a set of peer processes to elect a leader or a coordinator. It can be necessary to determine a new leader if the current one fails to respond. Provided that all processes have a unique identification number, leader election can be reduced to finding the non crashed process with the highest identifier.

### A. Original Bully Algorithm by Garcia Molina

Bully algorithm is one of the most famous election Algorithms which was proposed by Garcia-Molina [10] in 1982. It is briefly described in this section with its limitations.

#### 1. Algorithm

This algorithm is established on some basic assumptions which are:

➢ It is a synchronous system and it uses timeout mechanism to keep track of coordinator failure detection [11].
➢ Each process has a unique number to distinguish them [10, 13].
➢ Every process knows the process number of all other processes [10].
➢ Processes do not know which processes are currently up and which processes are currently down [8, 9, 10].
➢ In the election, a process with the highest process number is elected as a coordinator which is agreed by other alive processes [12].

is crashed because of message timeouts or failure of the coordinator to initiate a handshake, it executes bully election algorithm using the following sequence of actions (figure 1).

- ➢ P sends an election message (inquiry) to all other processes with higher process numbers respect to it. If P doesn't receive any message from processes with a higher process number than it, it wins the election and sends a coordinator message to all alive processes.
- ➢ If P gets answer message from a process with a higher process number, P gives up and waits to get coordinator message from any of the process with higher process number. Then new process initiates an election and sends election message to processes with higher process number than that one. In this way, all processes will give up the election except one which has the highest process number among all alive processes and it will be elected as a new coordinator. New Coordinator broadcasts itself as a coordinator to all alive processes in the system.
- ➢ Immediately after the recovery of the crashed process is up, it runs bully algorithm.

3. *Limitation*

Bully algorithm has following limitations:
- ➢ The main limitation of bully algorithm is the highest number of message passing during the election and it has order $O(n^2)$ which increases the network traffic.
- ➢ When any process that notices coordinator is down then holds a new election. As a result, there may n number of elections can be occurred in the system at a same time which imposes heavy network traffic.
- ➢ As there is no guarantee on message delivery, two processes may declare themselves as a coordinator at the same time. Say, p initiates an election and didn't get any reply message from Q, where Q has a higher process number than p. At that case, p will announce itself as a coordinator and as well as Q will also initiate new election and declare itself as a coordinator if there is no process having higher process number than Q.
- ➢ Again, if the coordinator is running unusually slowly (say system is not working properly for some reasons) or the link between a process and a coordinator is broken for some reasons, any other process may fail to detect the coordinator and initiates an election. But the coordinator is up, so in this case it is a redundant election.
- ➢ Again, if a process p with lower process number

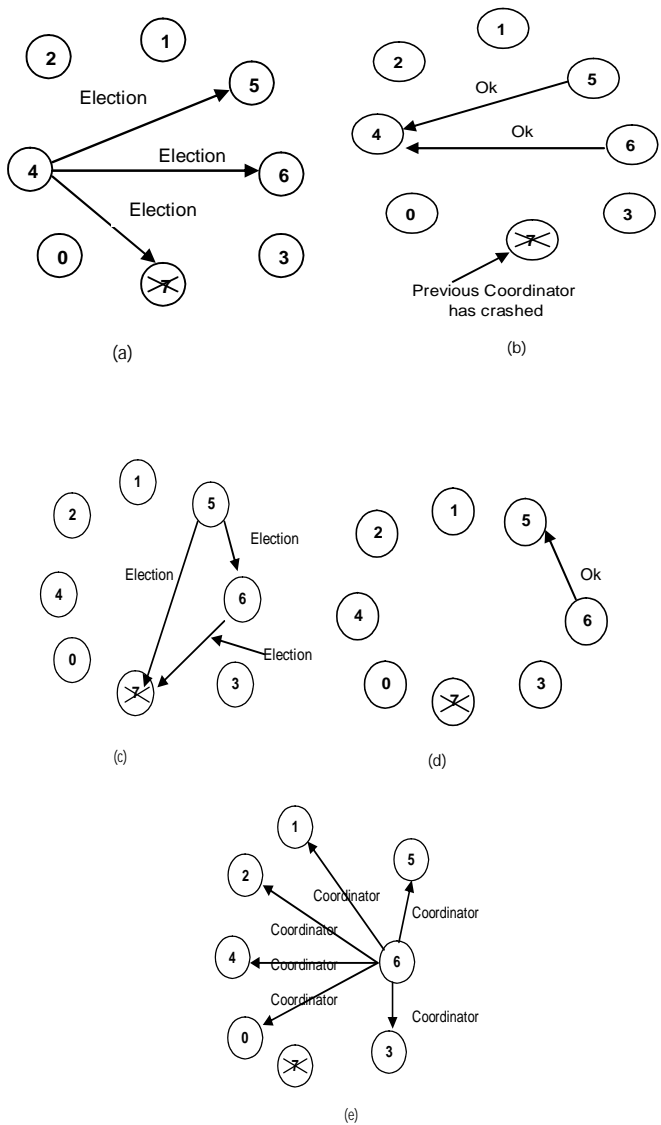

Fig.1. Original Bully Algorithm: (a) process 4 detects coordinator is failed and holds an election, (b) process 5 and 6 respond to 4 to stop election, (c) each of 5 and 6 holds election now, (d) process 6 responds to 5 to stop election, (e) process 6 winds and announces to all.

- ➢ A failed process can rejoin in the system after recovery [12].

In this algorithm, there are three types of message and there is an election message (inquiry) which is sent to announce an election, an answer (ok) message is sent as response to an election message and a coordinator (victory) message is sent to announce the new coordinator among all other alive processes [9].

2. *Procedure*

When a process P determines that the current coordinator than the current coordinator, crashes and recovers again, it will initiate an election where the current coordinator will win again. This is also a redundant election.

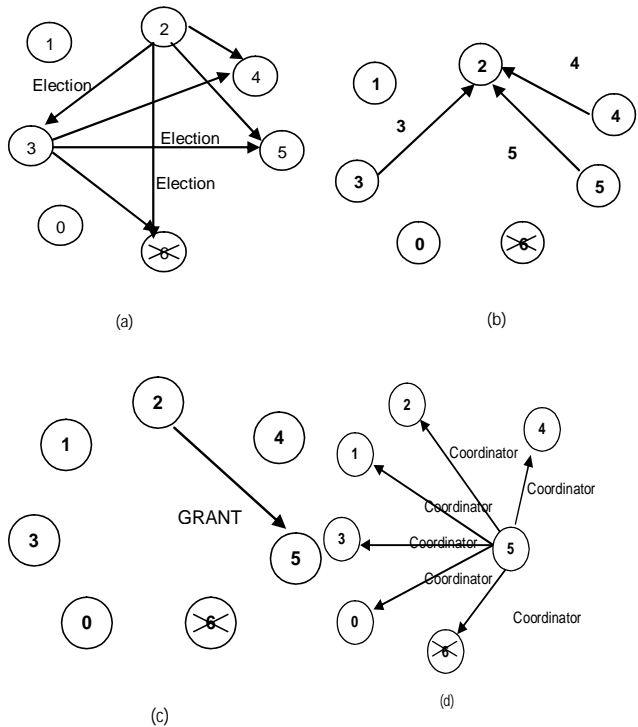

Fig.2. Modified Bully algorithm by M.S. Kordafshari et al : (a). process 2 detects coordinator is failed and holds an election, (b). process 3, 4 and 5 respond with their process number, (c). process 2 selects highest process number 5 and send a grant message to 5, (d). process 5 sends coordinator message to all processes.

### B. Modified Bully algorithm by M.S. Kordafshari et al.

M. S. Kordafshari et al. discussed the drawback of synchronous Garcia Molina's Bully Algorithm and modified it with an optimal message algorithm [8]. They showed that their algorithm is more efficient than Garcia Molina's Bully algorithm, because of fewer message passing and fewer stages.

#### 1. Algorithm

According to M. S. Kordafshari et al. [8], their proposed algorithm is briefly described below (figure 2).

➢ When process p notices that coordinator is down, it initiates an election by sending ELECTION message to all processes with higher priority number. If no process responses to p, it declares itself as a new coordinator. If some processes response to p, it will select the process with highest priority number and send back a GRANT message to that selected process. Finally selected process broadcast a coordinator message to all others as a coordinator itself. If any process with the highest priority number is up, it will run the algorithm again.

➢ To reduce concurrence election, when process p notices that the coordinator is down, it initializes election. If process q (q may be p) receives an ELECTION message from any process with lower priority number, it waits for a short time and replies to the process with lowest priority number and stop its own algorithm. But if p neither receives any response nor any ELECTION message from other processes with lower priority number, it declares itself as a coordinator.

#### 2. Limitation

Although this method reduces message passing complexity on some extend, it has following drawbacks.

➢ If a process p crashes after sending ELECTION message to higher processes or crashes after receiving priority number from higher processes, higher processes will wait for 3D (D is average propagation delay) time for coordinator broadcasting and if they don't receive any coordinator message, they will initiate modified algorithm again [8]. If there are q different higher processes, then there will be q different individual instance of modified algorithm at that moment in the system. Those are redundant election.

➢ If process p sends GRANT message to the process with the highest priority number, and p doesn't receive COORDINATOR message from that process with in D time, p will repeats the algorithm, which is redundant election. As after any process with higher priority number compare to coordinator is up, it runs the algorithm, it increases redundant elections.

➢ Although q sends stop message to p, if any other process r lower than q sends ELECTION message to q with this condition r<p<q, it takes network resources to send stop messages and increases network traffic.

➢ Every redundant election takes resources, increases total message passing and increases network traffics.

### C. Modified Bully algorithm by Quazi Ehsanul Kabir Mamun et al.

Quazi Ehsanul Kabir Mamun et al. described an efficient version Bully algorithm to minimize redundancy in electing the coordinator and to reduce the recovery problem of a crashed process [9].

#### 1. Algorithm

According to Quazi Ehsanul Kabir Mamun et al [9], their proposed algorithm is briefly described below.

➢ There are five types of message. An election message is sent to announce an election, an ok message is sent in response to an election message, on recovery, a process sends a query message to the processes with process number higher than it to know who the new coordinator is, a process gets an answer message from any process numbered higher than it in response to a query message and a coordinator message is sent to announce the number of the elected process as the new coordinator.
- When a process p notices that coordinator is down, it sends an election message to all processes with higher number. If no response, p will be the new coordinator. If p gets ok message, it will select the process with highest process number as coordinator and send a coordinator message to all process.
- When a crashed process recovers, it sends query message to all process with higher process number than it. And if it gets reply then it will know the coordinator and if it doesn't get any reply it will announce itself as a coordinator.

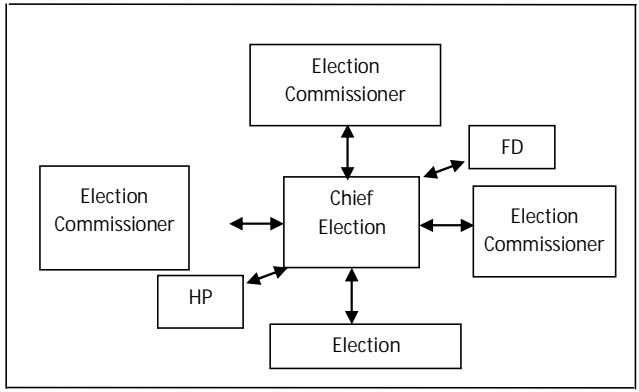

Fig.3. Architecture of Election Commission (EC)

*2. Limitation*

Although this algorithm reduces redundant election on some extent, it still has some redundant elections and also has high message complexity. Some of the limitations are given below:
- On recovery, it sends query message to all processes with higher process number than it, and all of them will send answer message if they alive. Which increases total number of message passing and hence it increases network traffic.
- It doesn't give guarantee that any process p will receive only one election message from processes with lower process number. As a result there may be q different processes with lower process number can send election message to p and p will send ok message to all of them. This increases number of election and also number of message passing.
- It doesn't give any idea if p will crash after sending an election message to all processes with higher process number.
- It also doesn't give any idea if a process with the highest process number will crash after sending ok message to p.

IV. METHODOLOGY

A. *Election Commission (EC)*

Election Commission is an electoral administrative body established to deal with leader election mechanism in a distributed computing system. It is constructed by a group of special processes in distributed system. It is authorized to handle the whole election process. It defines the rules and regulations for attending in an election process in a distributed computing system. It has one Chief Election Commissioner (CEC) and four Election Commissioners. If any of the commissioners failed, Election Commission will recover that commissioner immediately and other processes do not have concern of that. An Election Commission has a unique group ID. Other processes in the system communicate with Election Commission using this group ID. As a result, if any of the commissioners is down, there will be not any problem in election. It has a reliable failure detector (FD) [9]. If maximum message transmission delay is $T_{msg}$ and maximum message processing delay is $T_{pos}$ then maximum time required to get a reply after sending a message to any process from Election Commission is $T = 2T_{msg} + T_{pos}$ [9]. If Election Commission does not get any reply from a process within T time, then FD of Election Commission will report that requested process is down. As like as FD, Election Commission has another component named helper (HP), the function of HP is to find out the process with the highest process number using sending alive message. It knows process number of all processes of the system. Figure 3 represents the architecture of an EC.

B. *Chief Election Commissioner*

Chief Election Commissioner is the principal of Election Commission. The process with the highest priority in Election Commission will be the Chief Election Commissioner. It administrates other election commissioners and handles FD and HP.

C. *Election Commissioner*

Election Commissioner is a member of Election Commission. It is a special kind of process. Any Election Commission in a distributed system will have a few numbers of Election Commissioners (say four). All of them consult with the Chief Election Commissioner under the rules and regulation while there will be a need of an election to elect a coordinator in a distributed system.

## V. PROPOSED MODIFIED APPROACH FOR BULLY ALGORITHM

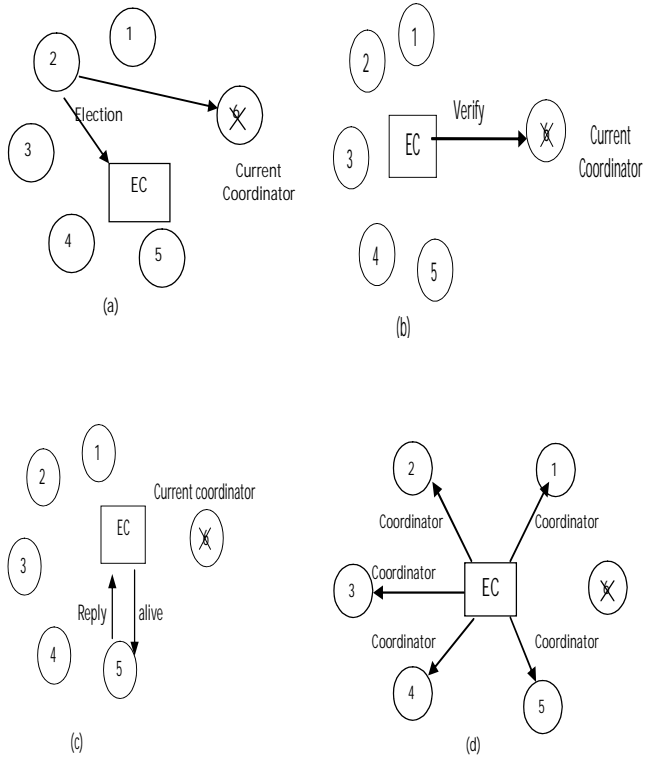

Fig.4. *Election Procedure:* (a) *Process 2 detects current coordinator is down and sends an election message to EC,* (b) *EC verifies either the coordinator is really down or not,*(c) *EC finds the alive process with highest number using alive message,*(d) *EC sends coordinator message to all process having process number of currently won.*

In Section 3, different version of modified bully algorithms [8, 9] and the limitations of original bully algorithm [10] have been clearly stated. It is clear that each algorithm has lot of redundant elections and message passing are also high between processes. Due to redundant election and high message passing, these algorithms impose heavy traffic in network.

As the system is synchronous and Election Commission has a FD and HP to solve limitations which is mentioned in section 3, we have proposed a modified version of bully algorithm using election commission concept. This algorithm not only reduces redundant elections but also reduces message passing between processes. And hence traffic in network will be decreased dramatically.

### A. Algorithm
Our proposed algorithm has the flowing steps:

- When process P notices that the coordinator is down, it sends an ELECTION message to Election Commission.
- FD of Election Commission verifies ELECTION message sent by P. If the sending notice of P is not correct, then Election Commission will send a COORDINATOR message to P with process number of the current coordinator.
- If the sending notice of P is correct and if the highest process number is P, then Election Commission will send a COORDINATOR message to all processes with process number of P as a new coordinator. If the highest process number is not P, Election Commission will simply find out the alive process with the highest process number using HP and sends a COORDINATOR message to all processes with the process number of that process as a new coordinator.
- If any process including last crashed coordinator is up, it will send a QUERY message to the Election Commission. If the process number of the newly entranced process is higher than the process number of the current coordinator, Election Commission will send a COORDINATOR message to all processes having the process number of new coordinator. It not, Election Commission will simply send a COORDINATOR message to newly entranced process having process number of the current coordinator.
- If more than one process sends ELECTION message to Election Commission at the same time, then Election Commission will consider the process with higher process number which ensure less message passing to find out the highest process number using HP.

### B. Procedure
Figure 4 represents regular election procedure of the proposed algorithm. Here, the system consists of six processes with process number 1 to 6. Current coordinator is the process 6. But it has just crashed and process 2 first notices this. So it sends an election message to the EC in Figure 4(a).In Figure4 (b), EC sends verify message to the current coordinator to be sure about the election message sent by process 2.

After verification, In Figure 4 (c), EC sends alive message to process 5 (the next highest process number) to check either the current highest process is alive or not. And EC gets a reply message from 5. In Figure 4 (d), EC select 5 as new coordinator and sends coordinator message to all processes having 5 as a new coordinator of the system.

Figure 5 represents the steps when a crashed process is up. In Figure 5 (a), last crashed coordinator 6 is up and sends a query message to EC. As process number of 6 is higher than the current coordinator of the system, in Figure

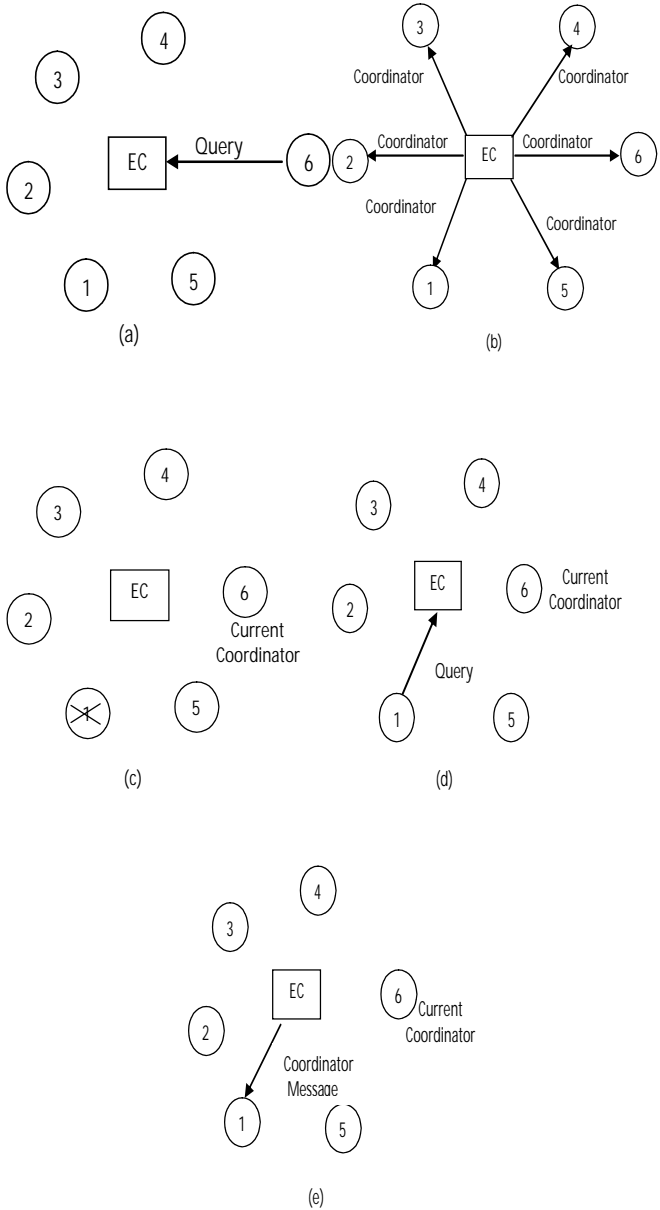

Fig.5. *Query after Recovery: (a) Last crashed coordinator 6 is up and sends a query message to the EC, (b) EC selects 6 as new coordinator and sends coordinator message to all processes, (c) Now process 1 is crashed, (d) Again process 1 is up and sends query message to EC,(e) EC sends coordinator message to process 1 having the current coordinator).*

5 (b), EC sends coordinator message to all processes with process number 6 as new coordinator. In figure 5 (c), process 1 is now just Crashed. In figure 5 (d), process 1 is just up after crashed, and it sends a query message to EC.

EC checks that process number of newly entranced is lower than the current coordinator. So in Figure 5 (e) EC sends coordinator message to only process 1 having the process number of current coordinator of the system.

At any time, if more than one processes notice that coordinator is down, they will send election message to EC. After verification, EC will consider election request of the process having higher process number. In Figure 6, process 4 and 5 detect that coordinator 6 is down, So 4 and 5 send election message to EC. After verification, EC only consider election message of process 5. It ensures less message passing to find out the highest process number.

Say if EC considers election message of 4, then according to our algorithm, EC will have to send alive message to 5 to find higher process number. But if EC considers election message of 5, it doesn't need to send alive message because, 5 is already the higher process number and EC can select 5 as new coordinator. This was EC can ensure less message passing.

## VI. COMPARISON AND DISCUSSION

In this section, we present the comparison in different issues among our proposed algorithm, original bully algorithm [10] and existing modified versions of bully algorithm [8, 9]. We consider message passing complexity and redundant election both of which increase network traffic.

### A. Message passing

If there are n processes in the system and p is the process number which detects failure of coordinator, then

- In original bully algorithm [10], there will be needed of   message passing between processes. In the worst case, if process with the lowest process number detects coordinator as failed, then it requires   message passing. In the best, case when p is the highest process number, it requires   messages.
- For the case of modified bully algorithm [8] there will be need of   or   messages passing between processes. In worst case that is the process with lowest process number detects coordinator as failed, it requires 3n-1 messages passing. In best case when p is the highest process number, it requires (n-p) + n messages.
- For the case of modified bully algorithm [9] there will be need of   or O(n) message passing between processes. In worst case that is the process with lowest process number detects coordinator as failed, it requires 3n-1 message passing. In best case when p is the highest process number, it requires (n-p) + n messages.

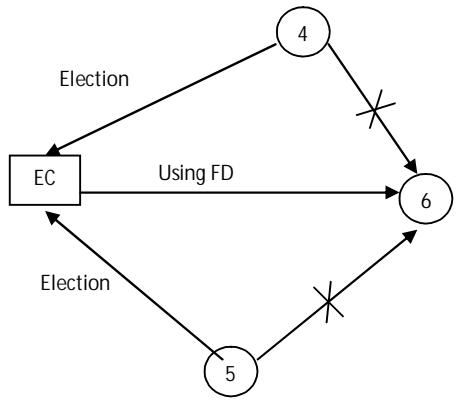

Fig.6. More than one election Message to EC.

- For the case of our proposed algorithm there will be need of 1 election message to inform EC, 2 verify message to ensure the failure of coordinator, and say r is the highest alive process then alive and reply message to find out the highest alive process and so total or O (n) message passing between processes. If the process with lowest process number detects coordinator as failed it will not change total message. In worst case it may happen that our algorithm needs to check up process to p+1 to find out highest alive process. Only at that case it requires message passing between processes. However, in best case, our algorithm may find the highest alive process with only one alive and one reply message that is highest alive process in the system is process with process number n-1. In that case, our algorithm requires only 1+2+2+n messages. When p is the highest process number, it requires only 1+2 + n messages.

- If a process crashes and recovers again, it sends a query message to all processes higher than that process [9] to know the current coordinator which requires 2*(n-p) message passing. But in our algorithm, any process after recovery will only send query message to EC and EC will send a coordinator message having process number of current coordinator which requires only 2 messages passing.

*B. Redundant election*

- In original bully algorithm [10] and modified bully algorithm [8, 9], if coordinator is running unusually slowly say (system is not working properly for some reasons) or the link between a process and coordinator is broken for some reasons, there will be redundant election, although current coordinator is up. But in our algorithm, as EC verifies either current coordinator is really up or down when EC receives any election message from any process, it ensures that there will be no redundant election in the system.

- If a process p crashes and recovers again, it initiates an election [8, 10] where the current coordinator wins again which is redundant election. But in our algorithm, after recovery, any process will send query message to EC and EC will reply with coordinator message having process number with current coordinator which reduces unnecessary election.

*C. Multiple Coordinators*

As there is no guarantee on message delivery that may happen more than one coordinator exist in the system at a time in bully algorithm [10] and modified bully algorithm [8, 9]. But in our algorithm there is no possibility of this as EC handles whole election process.

VII. CONCLUSION

In this paper, we modified bully algorithm using a new concept Election Commission (EC). We tried to overcome limitations of original bully algorithm [10] and modified bully algorithm [8. 9]. Our comparison and discussion section prove that our algorithm is more efficient than bully algorithm [10] and modified bully algorithm [8, 9] in respect of message passing, redundant election and network traffic.

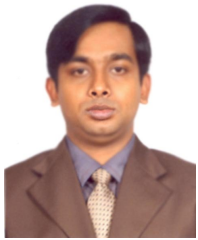
**Muhammad Mahbubur Rahman** received his master degree in Information Technology from Institute of Information Technology at University of Dhaka. Currently, he is working as a lecturer in the department of Computer Science at American International University-Bangladesh (AIUB). His research interests include Data Mining, Machine Learning, Bioinformatics, Game Theory, Artificial Intelligence, Econometrics and Logic Circuit Minimization. Currently his active research works are in bioinformatics and data mining. He has several years of working experience in local and international telecommunication and software companies.

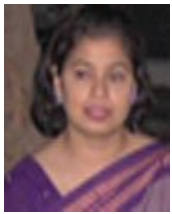
**Afroza Nahar** received her masters' degree in Computer Science from Asian Institute of Technology, Thailand and M. Sc in Applied Mathematics from University of Dhaka, Dhaka, Bangladesh. Currently, she is working as an Assistant Professor in the department of Computer Science and member of AIUB Quality Assurance Center (AQAC) at American International University-Bangladesh (AIUB). Her major research interests include Graph Theory, Linear Algebra, e-Commerce, e-Governance, Computational Mathematics, Knowledge Management, and Distributed Computing.